\documentclass[aps,prb,reprint,superscriptaddress,nofootinbib]{revtex4-2}
\usepackage{graphicx,epsfig}
\usepackage{times}
\usepackage{graphics,bm,float}
\usepackage{amssymb,amsmath,rotate,color,dsfont}
\usepackage[breaklinks,colorlinks=true,urlcolor=blue,citecolor=blue,linkcolor=blue]{hyperref}
\usepackage[T1]{fontenc}

\begin{document}
\unitlength 1 cm
\newcommand{\be}{\begin{equation}}
\newcommand{\ee}{\end{equation}}
\newcommand{\bearr}{\begin{eqnarray}}
\newcommand{\eearr}{\end{eqnarray}}
\newcommand{\nn}{\nonumber}
\newcommand{\vpdag}{{\vphantom{\dagger}}}
\newcommand{\vecr}{\vec{r}}
\newcommand{\bs}{\boldsymbol}
\newcommand{\up}{\uparrow}
\newcommand{\down}{\downarrow}
\newcommand{\fns}{\footnotesize}
\newcommand{\ns}{\normalsize}
\newcommand{\cdag}{c^{\dagger}}
\newcommand{\so}{\lambda_{\rm SO}}
\newcommand{\jh}{J_{\rm H}}
\newcommand{\sgn}{\text{sgn}}

\definecolor{red}{rgb}{1.0,0.0,0.0}
\definecolor{green}{rgb}{0.0,1.0,0.0}
\definecolor{blue}{rgb}{0.0,0.0,1.0}

\newcommand{\red}[1]{\textcolor{red}{#1}}
\newcommand{\violet}[1]{\textcolor{violet}{#1}}
\newcommand{\blue}[1]{\textcolor{blue}{#1}}

\title{Antiferromagnetic Chern insulator with large charge gap
in heavy transition-metal compounds
}

\author{Mohsen Hafez-Torbati}
\email{m.hafeztorbati@gmail.com}
\affiliation{Department of Physics, Shahid Beheshti University, 1983969411, Tehran, Iran}

\author{G\"otz S. Uhrig}
\email{goetz.uhrig@tu-dortmund.de}
\affiliation{Condensed Matter Theory, Department of Physics, TU Dortmund University, 44221 Dortmund, Germany}

\begin{abstract}
Despite the discovery of multiple intrinsic magnetic topological insulators in recent years
the observation of Chern insulators is still restricted to very
low temperatures due to the negligible charge gaps. Here, we uncover the potential
of heavy transition-metal compounds for realizing a
collinear antiferromagnetic Chern insulator (AFCI)  with a charge gap as large as 300 meV.
Our analysis relies on the Kane-Mele-Kondo model with
a ferromagnetic Hund coupling $J_{\rm H}$ between the spins of itinerant electrons and the localized
spins of size $S$.
We show that a spin-orbit coupling $\lambda_{\rm SO} \gtrsim 0.03t$, where $t$ is the nearest-neighbor hopping element,
is already large enough to stabilize an AFCI provided the
alternating sublattice potential $\delta$ is in the range
$\delta \approx SJ_{\rm H}$.
We establish a remarkable increase in the charge gap upon increasing $\lambda_{\rm SO}$ in the AFCI phase.
Using our results we explain the collinear AFCI recently found in monolayers of CrO and MoO with
charge gaps of 1 and $50$ meV, respectively. In addition,
we propose bilayers of heavy transition-metal oxides of perovskite structure as
candidates to realize a room-temperature AFCI if grown along the $[111]$ direction
and subjected to a perpendicular electric field.
\end{abstract}

\maketitle

\onecolumngrid

\section*{Introduction}

Magnetic topological insulators combining non-trivial electronic topology with magnetic properties
have attarcted a lot of attention in the past decade as they allow for the observation
of novel quantum phenomena such as the quantum anomalous Hall state also known as Chern insulator (CI) 
\cite{Tokura2019,Chang2023,Wang2023}.
A CI displays the quantum Hall effect at zero magnetic field \cite{Haldane1988} and is promising for the
construction of the next-generation electronic devices with dissipationless charge and spin transport.
The realization of the CI also opens a promising route to the detection
of the long thought chiral Majorana fermions as a quasiparticle in condensed matter physics.
These features justify the great interest in CIs from both fundamental and  applied
perspectives.

The realization of the CI requires strong spin-orbit coupling and 
spontaneous magnetic ordering.
The former grants non-trivial topology and the latter breaks the time-reversal symmetry permitting
a finite Chern number. The CI is realized in thin films of topological insulators doped with
magnetic impurities \cite{Chang2013,Kou2015,Mogi2015}, in thin films of the intrinsic magnetic topological insulator
MnBi$_2$Te$_4$ with an odd number of layers \cite{Deng2020},
in twisted bilayer graphene aligned to hexagonal boron nitride \cite{Serlin2020},
and in twisted transition metal dichalcogenide heterobilayers \cite{Li2021}. Despite the elimination of
the detrimental effect of disorder in the latter three systems the observation of
the quantum anomalous Hall effect is still restricted to low temperatures of a few Kelvin \cite{Deng2020,Serlin2020,Li2021}.
This is due to the small magnetic transition temperature and the negligible
charge gap. The magnetically induced charge gap in the top and the bottom ferromagnetic
layers of MnBi$_2$Te$_4$ is extremely small or almost absent \cite{Li2019b,Chen2019,Hao2019,Swatek2020}.
This limits the precise quantization of the anomalous Hall effect to a temperature one order of 
magnitude smaller than the magnetic transition temperature of the material \cite{Deng2020}.
To reach  realizations of the quantum anomalous Hall effect at higher temperatures, finding CIs in
new classes of systems with higher magnetic transition temperatures and sizable charge gaps is indispensable.

Antiferromagnets represent a large class of materials with unique properties which are likely to
shape the future of the spintronic and magnonic 
technology operating in the terahertz regime \cite{Jungwirth2016,Baltz2018,malki20b,barma21}.
They display generally a higher magnetic transition temperature in contrast to their ferromagnetic counterparts.
They allow for more densely packed domains serving as bits because they do not display long-range
stray fields since they are lacking a net magnetization averaged over all sites \cite{loth12}.
Furthermore, the generic energies in antiferromagnets are higher by roughly three orders
of magnitude so that one can hope for faster data manipulation \cite{gomon14,kampf11}.

Collinear antiferromagnetic (AF) ordering in Mott and charge-transfer insulator materials are
reported to be accompanied by a strong blue shift of the charge gap \cite{Ferrer-Roca2000,Bossini2020}. The observed strong magnetic
blue shift stems from the fact that the electrons defining the charge gap are the ones
responsible for the magnetic ordering \cite{Bossini2020}. This implies that the magnetic ordering has
a direct effect on the charge gap as is nicely described based on the exchange and the double exchange mechanisms
in a Kondo lattice model for the electrons in $d$ orbitals \cite{Hafez-Torbati2021,Hafez-Torbati2022}.
This behavior must be contrasted to the one in a system such as MnBi$_2$Te$_4$ where
the charge gap is defined by the band gap of weakly interacting electrons in $s$ and $p$ orbitals
while the magnetic ordering is due to electrons in partially filled $d$ orbitals \cite{Li2019b}.

With these motivations we search in this paper for a CI phase with collinear AF order (AFCI) 
in a Kondo lattice systems which is described by the Kane-Mele-Kondo (KMK) model
on the honeycomb lattice at half-filling.
A particular focus is the magnitude of the charge gap which we want to be sizable
to warrant robustness and observability at elevated temperatures.
A ferromagnetic Kondo interaction (Hund coupling $\jh>0$) couples
the spins of itinerant electrons to localized spins with spin quantum number $S$.
We recall that model Hamiltonians, for which reliable results can be obtained, played a paramount role in the
field of topological insulators \cite{Haldane2017}. In analogy, it is also our strategy here to
search for an AFCI phase with a large charge gap
on the level of model Hamiltonians leaving the identification of specific compounds to future research.

For a fixed value of the spin-orbit coupling $\lambda_{\rm SO}$
in units of the nearest-neighbor (NN) hopping parameter $t$ 
we map out the phase diagram of the KMK model as function of
the rescaled Hund coupling $SJ_{\rm H}$ and the alternating sublattice potential $\delta$;
recall that the honeycomb lattice is bipartite.
We will provide firm evidence for the emergence of a collinear AFCI phase in the region $\delta \approx S\jh$.
A spin-orbit coupling $\so \gtrsim 0.03t$ is already large enough to stabilize the AFCI phase.
These findings signal the possibility to realize the AFCI even in $3d$ materials
which display a relatively weak spin-orbit coupling.
The charge gap $\Delta$ in the AFCI phase
increases remarkably upon increasing the spin-orbit coupling $\lambda_{\rm SO}$
described by $\Delta \approx 14t (\lambda_{\rm SO}/t-0.03)$  up to $\so \approx 0.15t$.
It saturates to $\Delta \approx 2.5t$ for large values of  the spin-orbit coupling $\lambda_{\rm SO}$.

Based on these findings we discuss the collinear AFCI
recently found in first-principle calculations for monolayers
of CrO  with a charge gap of 1 meV \cite{Guo2023}
and for monolayers of MoO with the charge gap of 50 meV \cite{Wu2023}.
Our estimates suggest the possibility of the realization of an AFCI with a charge gap as large as 300 meV in
$5d$ transition-metal compounds. Additionally,
we propose bilayers of heavy transition-metal oxides with perovskite structure
as candidates to realize a large-gap AFCI which can persist up to the room temperature.
These layers should be grown along the $[111]$ crystallographic axis and be
subjected to a perpendicular electric field to tune the alternating sublattice potential $\delta$ \cite{Xiao2011}.

\section*{Results}

\noindent{\bf Model Hamiltonian.}
The KMK model is given by
\begin{align}
H=&+t\sum_{\langle i,j\rangle} \sum_{\alpha} c^{\dag}_{i\alpha} c^{\vpdag}_{j\alpha}
+{\rm i}\lambda_{\rm SO} \sum_{[ i,j ]}
\sum_{\alpha \beta}
\nu_{ij}^{\vpdag}
c^{\dag}_{i\alpha} \sigma^{z}_{\alpha\beta} c^{\vpdag}_{j\beta} \nn \\
&+\sum_{i} \sum_{\alpha} \delta_i^{\vpdag}  c^{\dag}_{i\alpha} c^{\vpdag}_{i\alpha}
-2J_{\rm H} \sum_{i} \vec{s}_i \cdot \vec{S}_i \ ,
\label{eq:kmk}
\end{align}
where $c^{\dag}_{i\alpha}$ and $c^{\vpdag}_{i\alpha}$ are the  creation and annihilation fermion operators
at the site $i$ on the honeycomb lattice with the $z$-component of its spin $\alpha=\uparrow$ or $\downarrow$.
The first term in Eq.\ \eqref{eq:kmk} is the nearest-neighbor (NN) hopping where
 $\langle i,j\rangle$ denotes adjacent sites $i$ and $j$.
The second term originates from the spin-orbit coupling assuming that the honeycomb lattice lies in the
 $xy$ plane.
The notation $\left[ i,j\right]$ denotes next-nearest-neighbor sites $i$ and $j$,
$\sigma^{z}$ is the corresponding Pauli matrix and
$\nu_{ij}=2/\sqrt{3}(\hat{d}_1 \times \hat{d}_2)_{z}=\pm 1$ where
$\hat{d}_1$ and $\hat{d}_2$ are the unit vectors along the two bonds the electron passes from site $j$
to site $i$. We assume $\so \geq 0$ without the loss of generality.
The third term is the alternating sublattice  potential with
the on-site energy $+\delta$ on sublattice $A$ and $-\delta$ on sublattice $B$.

We emphasize that this alternating potential can be fine-tuned in buckled structures
using a perpendicular electric field \cite{Ezawa2012,Xiao2011}. So it is a realistic
knob to control the system's properties.
Without a finite alternating potential the effect of the time-reversal transformation
can be undone by a space group transformation (here: mirror reflection) 
which prevents the emergence of an AFCI 
\cite{Ebrahimkhas2022,Jiang2018,Ebrahimkhas2021}. Thus, a finite value of $\delta$ 
is crucial, see below.
Applying strain is another means to break such a composite symmetry so that an AFCI
can emerge \cite{Guo2023,Wu2023}. Then, the strain can be used to 
fine-tune the model parameters \cite{Yan2015,Liu2014}.
The first three terms in Eq.\ \eqref{eq:kmk} define the Kane-Mele model $H_{\rm KM}$ \cite{Kane2005b,Kane2005a}
and the last term represents the ferromagnetic Kondo interaction $H_{\rm K}$
between the spin of the itinerant electron at site $i$
$\vec{s}_i$ and the localized spin $\vec{S}_i$. Its value is controlled by the Hund coupling $J_{\rm H}>0$.
The size of the localized spin $\vec{S}_i$ is denoted by $S$.

The Kane-Mele model for $\delta<3\sqrt{3}\lambda_{\rm SO}$ 
represents a quantum spin Hall insulator (QSHI) with the opposite spin components
having the opposite Chern numbers, $\mathcal{C}_{\uparrow}=+1$ and $\mathcal{C}_{\downarrow}=-1$.
For simplicity, we henceforth assume $\mathcal{C}_{\alpha}=\sgn(\alpha)$
where $\sgn(\up)=1$ and $\sgn(\down)=-1$.
This state is characterized by the $\mathds{Z}_2$ topological invariant
$\mathcal{C}_s=(\mathcal{C}_{\uparrow}-\mathcal{C}_{\downarrow})/2=1$.
At $\delta=3\sqrt{3}\lambda_{\rm SO}$ the charge gap closes and the transition between the QSHI and the
trivial band insulator with $\mathcal{C}_{\alpha}=0$ takes place.
An infinitesimal Kondo interaction couples the
spins of the itinerant electrons and the localized spins and leads 
to magnetic ordering induced by the van Vleck mechanism
\cite{Vleck1932,Bloembergen1955}.

In the strong coupling limit $(SJ_{\rm H}- \delta) \gg t,\so$, the subspaces with a finite number of double
occupancies are high in energy and can be rotated away perturbatively
by a unitary transformation \cite{Takahashi1977,Fazekas1999}
leading to the effective low-energy spin model
\be
\label{eq:spinm}
H_{\rm eff}=J_1\sum_{\langle i,j \rangle} \vec{s}_{i} \cdot \vec{s}_{j}
+J_2 \sum_{\left[ i,j \right]} ( {s}^z_{i} {s}^z_{j}\!-\! {s}^x_{i} {s}^x_{j}\!-\!{s}^y_{i} {s}^y_{j})
+H_{\rm K} \ ,
\ee
with $J_1={SJ_{\rm H}t^2}/(S^2J_{\rm H}^2-\delta^2)$ and $J_2=\lambda_{\rm SO}^2/(SJ_{\rm H})$.
The first term stabilizes the AF N\'eel order. The second term originates from the spin-orbit coupling
and reduces the SU(2) symmetry to a U(1) symmetry. This term favors that the local magnetic moments 
lies in the $xy$ plane (abbreviated $xy$-AF phase) rather than in the $z$ direction (abbreviated $z$-AF phase).
Within the mean-field approximation of the Hamiltonian \eqref{eq:spinm}
the energy difference per lattice site between the $z$-AF phase and the $xy$-AF phase 
is given by
\be
\label{eq:ediff}
\varepsilon_z - \varepsilon_{xy}=3J_2={3\so^2}/(S\jh) \ . 
\ee
This already illustrates the importance of considering the $xy$-AF solution in any analysis of the KMK model.

The third interesting limit of the KMK model \eqref{eq:kmk} is the atomic limit $t=\so=0$.
For $\delta>S\jh$, the itinerant electrons occupy the sublattice $B$ leaving the sublattice $A$
empty. For $\delta<S\jh$, there is exactly one itinerant electron at each lattice site. These two 
configurations with completely different charge distributions become degenerate at $\delta=S\jh$.
Introducing the hopping and the spin-orbit coupling allows for charge fluctuations and exotic
phases in the region $\delta {\approx }S\jh$ can be expected.

\begin{figure*}[t]
   \begin{center}
   \includegraphics[width=0.94\textwidth,angle=0]{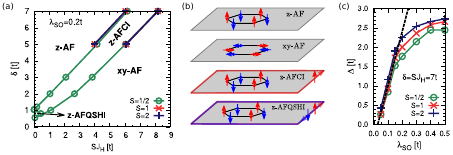}
   \caption{{\bf Phase diagram and charge gap.} (a) Phase diagram of the Kane-Mele-Kondo model \eqref{eq:kmk} for spin-orbit coupling
   $\so=0.2t$. There always exists an AF N\'eel order except for vanishing Hund coupling $\jh \equiv 0$.
	The local magnetic moment is oriented
   perpendicular to the honeycomb plane in the $z$-AF phase and in the honeycomb plane in the $xy$-AF
   phase. The non-trivial topological phases are the $z$-AF Chern insulator ($z$-AFCI) and the $z$-AF quantum
   spin Hall insulator ($z$-AFQSHI). In addition to the results for localized spins $S=1/2$ the phase diagram
   also includes data for $S=1$ and $2$ for selected parameters.
	 (b) Schematic representation of the different phases. 
	 The red and the blue filled circles represent lattice sites with different onsite energies. Such an
energy difference prevents the effect of the time-reversal transformation to be
compensated by a space group operation and allows for the existence of the AFCI with spin-polarized edge modes.
   (c) Charge gap $\Delta$ vs.\ $\so$ in the
   $z$-AFCI at $S\jh=\delta=7t$ for different  $S$. The black dashed line denotes a linear fit 
	to the data for $\so\leq 0.15t$ according to Eq.\ \eqref{eq:gap}.
	The results are obtained using DMFT with an ED
   impurity solver with $n_b=5$ bath sites.}
   \label{fig:main}
   \end{center}
\end{figure*}

{\bf Phase diagram and charge gap.}
We employ dynamical mean-field theory (DMFT) \cite{Georges1996} implemented in real space with 
the exact diagonalization (ED) impurity solver \cite{Caffarel1994}
to explore the bulk as well as the edge properties of the KMK model.
For the topological invariants of the bulk, we resort to
the topological Hamiltonian  \cite{Wang2012}, 
see ``Method'' for details.
Fig.\ \ref{fig:main}(a) displays the phase diagram of the KMK model
as function of the rescaled Hund coupling $S\jh$
and the alternating potential $\delta$ for spin-orbit coupling $\so=0.2t$.
The different phases are sketched schematically in Fig.\ \ref{fig:main}(b).
There is AF N\'eel order for any $\jh \neq 0$.
The topologically non-trivial $z$-AFCI emerges around
$S\jh \approx \delta$ separating the trivial $z$-AF phase at weak 
Hund coupling and the trivial $xy$-AF phase at strong Hund coupling.

The existence of the $z$-AFCI is demonstrated not only by computing the 
topological invariant using the topological 
Hamiltonian but also by showing the existence of gapless edge modes
for only one spin component in stripes of the interacting model \eqref{eq:kmk}.
In addition to the $S=1/2$ case,
data are provided for $S=1$ and $2$ for selected parameters to support
that the $z$-AFCI exists independent of the size of the localized spin.
We also find  a $z$-AFQSHI in the phase diagram Fig.\ \ref{fig:main}(a), but in a very 
small region so that we conclude that its stability beyond the DMFT remains
to be investigated.

In Fig.\ \ref{fig:main}(c) the charge gap $\Delta$ is plotted vs.\ the
spin-orbit coupling $\so$ in the $z$-AFCI phase for $S\jh=\delta=7t$. There is a 
remarkable increase in $\Delta$ in Fig.\ \ref{fig:main}(c)
upon increasing $\so$ up to $\so \approx 0.15t$. For even larger values of the spin-orbit
coupling the charge gap saturates at about $\Delta\approx 2.5t$.
The behavior is generically independent of the
size of the localized spin $S$ although we find a slightly larger gap for a larger $S$.
Performing a linear fit to the data for $\so \leq 0.15t$ we find
\be
\label{eq:gap}
\Delta \approx 14t (\so/t - 0.03) \ ,
\ee
shown by a black dashed line in Fig.\ \ref{fig:main}(c).

In most experimental realizations, the spin-orbit coupling $\so$ of Kane-Mele-type is not
very large and satisfies the relation
$\so \leq 0.15t$ so that Eq.\ \eqref{eq:gap} provides a ready-to-use formula to estimate the charge gap in
the AFCI phase.
 Since only a small spin-orbit coupling $\so > 0.03t$ is required to stabilize the AFCI phase
 the AFCI can even occur in $3d$ transition-metal compounds,
although with a rather small charge gap relative to the one in $4d$ and $5d$ systems.
Using Eq.\ \eqref{eq:gap} we will discuss the collinear AFCI occurring in recent first-principle
calculations for monolayers of CrO with a charge gap of $1$ meV \cite{Guo2023} and for monolayers of 
MoO with a charge gap of $50$ meV \cite{Wu2023}, see below.
In addition, we will unveil the possibility to realize
an AFCI with a charge gap as large as 300 meV in $5d$ materials.

\begin{figure}[t]
   \begin{center}
   \includegraphics[width=0.28\textwidth,angle=-90]{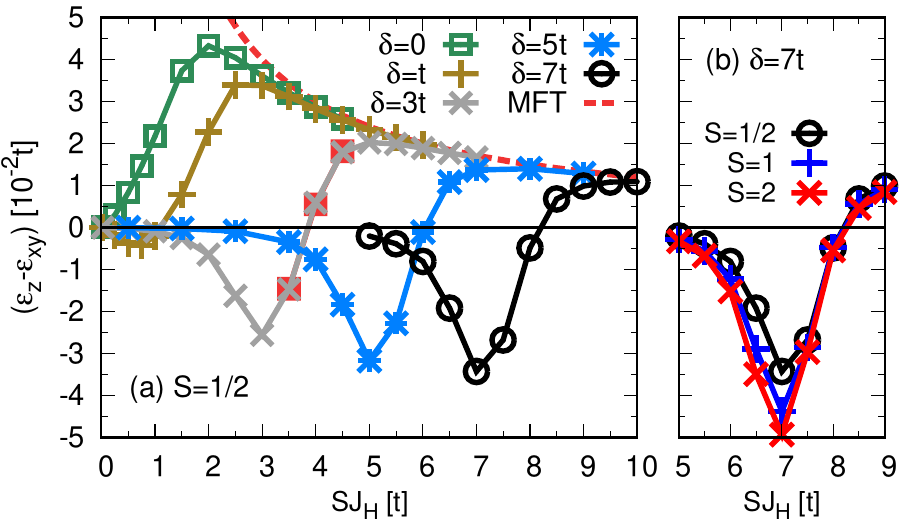}
   \caption{{\bf Spin-flop transition.} Ground state energy difference per lattice site between the $z$-AF state and the $xy$-AF state
   vs.\ the rescaled Hund coupling $S\jh$ for the spin-orbit coupling $\so=0.2t$. 
   The results in panel (a) are for fixed localized spin $S=1/2$ 
   and various values of the alternating sublattice potential $\delta$. The results in panel (b) are for fixed 
   $\delta=7t$ and different sizes of $S$. The data are obtained by DMFT 
	with ED impurity solver for $n_b=5$ bath sites except for the red filled squares at $\delta=3t$ in panel (a) 
	where $n_b=6$. The red dashed line in panel (a) represents the results of the static mean-field theory (MFT) 
	\eqref{eq:ediff} of the corresponding low-energy spin model \eqref{eq:spinm}.}
   \label{fig:energy}
   \end{center}
\end{figure}

{\bf Spin-flop transition.} First, we study the spin-flop transition from the $z$-AF order
to the $xy$-AF order ignoring the possibility of an intermediate phase.
In a second step, we address the phase transitions to and from the intermediate topological $z$-AFCI.
The energy difference per lattice site between the $z$-AF solution and the
$xy$-AF solution is plotted vs.\ the rescaled Hund coupling $S\jh$ in Fig.\ \ref{fig:energy} at 
spin-orbit coupling $\so=0.2t$.
The results in panel Fig.\ \ref{fig:energy}(a) are for localized spin $S=1/2$ and various
values of the sublattice potential $\delta$. Panel (b) compares the results for different values of 
$S$ at fixed value of the sublattice potential $\delta=7t$.
We computed multiple $xy$-AF solutions of the DMFT equations to check that they all have the same energy
and the same magnitude of the local magnetization. They differ in their orientation
in the $xy$ spin plane as usual for a spontaneous breaking of this U(1) symmetry.
The data are obtained for  $n_b=5$ bath sites except for the red filled squares 
at $\delta=3t$ in panel (a) which are evaluated for $n_b=6$. The perfect agreement
indicates that the results do not change significantly if the number of bath sites is increased.

Fig.\ \ref{fig:energy}(a) shows that for $\delta=0$ the $xy$-AF solution has an energy lower 
than the $z$-AF solution for any $\jh>0$.
This remains true also for finite, but small values of $\delta\lesssim 0.5t$. 
For larger values of $\delta$ and  small values of  $\jh$,
the $z$-AF state is energetically more favorable than the $xy$-AF state.
The $z$-AF state becomes most stable relative to the $xy$-AF state near $\delta=S\jh$.
Upon increasing $\jh$ the energy difference $\varepsilon_z-\varepsilon_{xy}$ crosses
the zero-energy line and a spin-flop transition to the $xy$-AF state takes place.
The transition occurs almost at the same value $S\jh$ for different sizes of the localized spin $S$
as can be seen from the results in Fig.\ \ref{fig:energy}(b).
A similar spin-flop transition has been observed in single-orbital models although there
the $z$-AF state exists only in a very narrow region \cite{Jiang2018,Ebrahimkhas2022}.

For large Hund couplings the energy difference between the $z$-AF solution and the
$xy$-AF solution approaches the mean-field theory value \eqref{eq:ediff} 
of the corresponding low-energy spin model depicted
by a red dashed line in Fig.\ \ref{fig:energy}(a) independent of the value of $\delta$.
This is expected because the local charge fluctuations, taken into account
in the DMFT of the KMK model, are already incorporated
in the low-energy spin model \cite{Hafez-Torbati2022}. Both approaches, however,
the DMFT of the KMK model and the mean-field theory
of the  low-energy spin model, neglect \emph{non-local} quantum fluctuations.
We point out that the charge degrees of freedom start to contribute to the low-energy properties
for smaller values of $\jh$. Then, the effective spin model \eqref{eq:spinm} becomes invalid.

\begin{figure}[hbt]
   \begin{center}
   \includegraphics[width=0.28\textwidth,angle=-90]{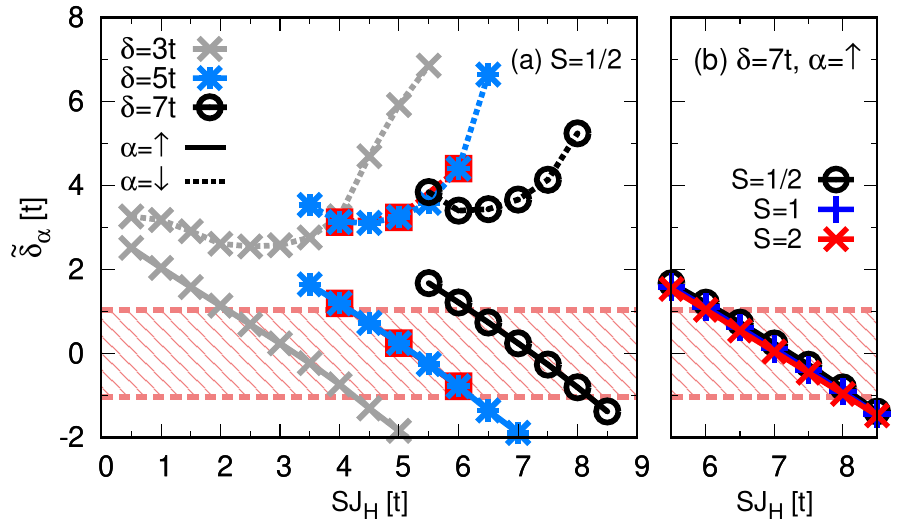}
   \caption{{\bf Effective alternating sublattice potential.}
   Effective alternating sublattice potential \eqref{eq:edelta} vs.\ the rescaled Hund coupling
	$S\jh$ for the spin-orbit coupling $\so=0.2t$. The results in panel (a) are for fixed localized spin $S=1/2$ 
	and various values of the bare alternating sublattice potential $\delta$. The results in panel (b) are for fixed $\delta=7t$ and various values of $S$. 
	The results refer to the $z$-AF solution
	where the local magnetic moment $\langle s_{i}^z \rangle$ is positive at the energetically higher sublattice $A$ 
	and negative at the energetically lower sublattice $B$.
   The data are obtained using DMFT with the ED impurity solver for $n_b=5$  bath sites 
   except for the red filled squares at $\delta=5t$ in panel (a)  which are for $n_b=7$. 
	The red dashed lines indicate the boundaries of the topological region (shaded area) present for
	$|\tilde{\delta}_\alpha|<3\sqrt{3}\so$.}
   \label{fig:edelta}
   \end{center}
\end{figure}

{\bf Topological phase transitions.}
The method using the topological Hamiltonian relates the topological invariant of an
interacting system to the one of an effective non-interacting model known as topological Hamiltonian \cite{Wang2012}
which consists of the bilinear part of the original Hamiltonian plus the self-energy at zero frequency,
see ``Method'' section.
Although the method has some limitations and
has to be used with caution \cite{Gurarie2011,Yoshida2014,He2016a,He2016,hawas23}
it can be applied to the different magnetic phases in our model. 
We stress that we do not rely solely on the topological Hamiltonian to characterize the AFCI.
We demonstrate below directly for the interacting KMK model \eqref{eq:kmk}
that gapless edge modes exist for only one spin component which is a smoking gun for
a topological non-trivial AFCI. Furthermore, we show for
the KMK model that the bulk charge gap closes at the topological phase transition.

In the $xy$-AF phase, the topological Hamiltonian corresponds to the Kane-Mele model subjected to an 
alternating in-plane magnetic field, see ``Method'' section. Similar to the uniform parallel magnetic field 
originally discussed  by Kane and Mele \cite{Kane2005b} an infinitesimal alternating in-plane magnetic 
field gaps the edge states and the system becomes a trivial band insulator. 
Hence, the $xy$-AF phase is always topologically trivial. 

In the $z$-AF state, the topological Hamiltonian corresponds to the Kane-Mele model 
with an effective spin-dependent sublattice potential resulting from the self-energy
\be
\label{eq:edelta}
\tilde{\delta}_\alpha=\delta+\Sigma_{A,\alpha}(0)
\ee
where $\Sigma_{A,\alpha}(0)$ is the zero-frequency self-energy on sublattice $A$ for spin component $\alpha$.
Thus, the spin component $\alpha$ is topological with  Chern number $\mathcal{C}_\alpha
=\sgn(\alpha)$ for $|\tilde{\delta}_\alpha|<3\sqrt{3}\so$; it is trivial with 
Chern number $\mathcal{C}_\alpha=0$ for $|\tilde{\delta}_\alpha|>3\sqrt{3}\so$. 
The spin-dependent sublattice potential opens the possibility that  one spin component is topological 
while the other spin component is trivial. This leads to the 
 $z$-AFCI with the non-zero Chern number
$\mathcal{C}=\mathcal{C}_\uparrow+\mathcal{C}_\downarrow\neq 0$.
To uncover the realization of this pivotal mechanism is one of the key results of this paper.

To characterize the topology of the $z$-AF state the effective sublattice potential \eqref{eq:edelta}
is displayed in Fig.\ \ref{fig:edelta} as function of the rescaled Hund coupling $S\jh$ for $\so=0.2t$.
Panel (a) shows results for $S=1/2$ and various values of the alternating sublattice potential $\delta$ 
while panel (b) compares the results for different sizes of the
localized spin $S$ at fixed sublattice potential $\delta=7t$.
The $z$-AF state is two-fold degenerate due to the spontaneous breaking of the time-reversal
symmetry. Here and in the following we always present results for the solution
where the local magnetic moment $\langle s_{i}^z \rangle$ is positive at the energetically higher sublattice $A$ 
and negative at the energetically lower sublattice $B$.
The red dashed lines in Fig.\ \ref{fig:edelta} mark the boundaries of the red shaded 
topological region $|\tilde{\delta}_{\sigma}|<3\sqrt{3} \so$. In the trivial region
$|\tilde{\delta}_{\sigma}|>3\sqrt{3} \so$ holds.
The data are for $n_b=5$  bath sites in the ED impurity solver except 
for the red filled squares at $\delta=5t$ in panel (a) where $n_b=7$.
The perfect agreement corroborates that chosen number of bath sites is sufficiently large.
In a static mean-field approximation of Eq. \eqref{eq:kmk} for the $z$-AF solution the $z$-AFCI appears for
$|\delta-S\jh|<3\sqrt{3}\so$ provided $|\delta+S\jh|>3\sqrt{3}\so$. This nicely matches
with the DMFT results in Fig.\ \ref{fig:edelta} and indicates that the quantum fluctuations are
negligible.

For small values of $S\jh$, see Fig.\ \ref{fig:edelta}(a),
both spin components are topologically trivial 
with $\mathcal{C}_\alpha=0$. As $S\jh$ is increased, one spin component
(spin $\uparrow$ in the figure) enters the topological region while
the other spin component remains topologically trivial implying a finite Chern number 
$\mathcal{C}=\mathcal{C}_\uparrow+\mathcal{C}_\downarrow \neq 0$ so that the system forms a $z$-AFCI.
The transition point shifts to larger values of $S\jh$ upon increasing $\delta$ in Fig.\ \ref{fig:edelta}(a).
The spin size has almost no effect, see Fig.\ \ref{fig:edelta}(b).
Increasing $S\jh$ further makes the spin $\up$ electrons leave the topological region so that
the $z$-AFCI state becomes topologically trivial again, i.e., the system enters the $z$-AF phase.
Comparing the results for $\delta=3t$ in Fig.\ \ref{fig:energy}(a) and in
Fig.\ \ref{fig:edelta}(a) manifests that the spin-flop transition to the $xy$-AF phase takes place 
\emph{before} the $z$-AFCI becomes $z$-AF insulator. This means that the $z$-AFCI undergoes
a phase transition to the $xy$-AF phase instead of entering the $z$-AF phase upon increasing $S\jh$.
For large values of $\delta$ the point where the $z$-AFCI becomes topologically trivial and
the point where the spin-flop transition to the $xy$-AF phase occurs almost coincide, see
the results for $\delta=7t$ in Fig.\ \ref{fig:energy}(a) and Fig.\ \ref{fig:edelta}(a).

Fig.\ \ref{fig:edelta}(a) shows results for intermediate to large sublattice potentials.
For small sublattice potentials $\delta<3\sqrt{3}\so \approx 1.04t $ both spin components can fall 
in the topological region with $\mathcal{C}_\alpha={\rm sgn}(\alpha)$ 
provided the Hund coupling $\jh$ is small enough.
Then, the system forms a $z$-AFQSHI with the $\mathbb{Z}_2$ topological invariant $C_s=1$. 
However, we find that such a phase is restricted to a very narrow region due to the spin-flop 
transition to the $xy$-AF phase as can be seen from Fig.\ \ref{fig:main}(a).  

For the emergence of the AFCI a finite alternating sublattice potential is essential because it prevents
that the effect of time reversal can be undone by a
space group operation. This lower symmetry is necessary for a finite Chern number 
\cite{Guo2023,Ebrahimkhas2022,Jiang2018,Ebrahimkhas2021}.
In contrast, the $z$-AFQSHI could, in principle, exist also at $\delta=0$. There is no $z$-AFQSHI at $\delta=0$
in Fig.\ \ref{fig:main}(a) due to the instability towards the $xy$-AF state. 
Finding a new class of systems which hosts the $z$-AFQSHI in a wide parameter region remains an intriguing 
open problem left to future research.

{\bf AFCI beyond the topological Hamiltonian.}
Gapless edge modes for only one spin component are a characteristic feature of the CI phase.
To confirm the existence of the AFCI beyond the topological Hamiltonian method
we compute the single-particle spectral function $A_{x,\alpha}(\omega)$ resolved in
spin $\alpha$ for the interacting KMK model \eqref{eq:kmk} for various values of $x$. 
No $y$ value needs to be specified because we use periodic boundary conditions in $y$ direction. 
The considered sample is a cylinder of  size
$L_x\times L_y=41\times 40$ with the edges at $x=0$ and $x=40$. For details, see 
Fig.\ \ref{fig:lattice} in the Method section.
The spectral function $A_{x,\alpha}(\omega)$ is averaged over the two inequivalent sites
\be
\label{eq:sf}
A_{x,\alpha}(\omega)= \left(A_{x,y;\alpha}(\omega)+A_{x,y+1;\alpha}(\omega)\right)/2
\ee
where $A_{x,y;\alpha}(\omega)$ is the spectral function at site $(x,y)$ for spin $\alpha$.
The number of bath sites is $n_b=5$ and a Lorentzian broadening of $0.05t$ is used.

\begin{figure}[t]
   \begin{center}
   \includegraphics[width=0.43\textwidth,angle=-90]{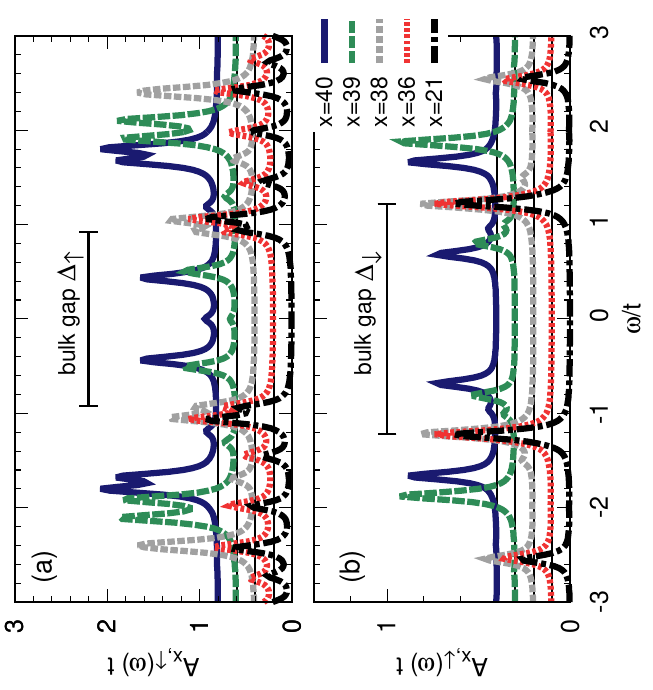}
   \caption{{\bf Gapless edge modes in AFCI.} Single-particle spectral function \eqref{eq:sf} near the Fermi energy $\omega=0$
   for spin $\up$ in panel (a) and spin $\down$ in panel (b) of a stripe of the
	size $L_x\times L_y=41\times 40$ with
   the edges at $x=0$ and $x=40$ and periodic boundary conditions in $y$ direction, i.e., a cylinder
	along the $x$ axis.
   The results are obtained for the KMK model \eqref{eq:kmk} with $\so=0.2t$, $S=1/2$, $\delta=7t$, 
	and $S\jh=7.25t$  so that the system is in the $z$-AFCI phase.
	The results correspond to the $z$-AF solution with the local magnetic moment of itinerant electrons
	$\langle s^z_i \rangle$ positive on the higher-energy sublattice $A$ and negative on the lower-energy
	sublattice $B$. The bulk gaps $\Delta_\alpha$ are indicated by the horizontal bars for comparison.
 The data are computed for  $n_b=5$ bath sites in the ED impurity solver.}
   \label{fig:sf}
   \end{center}
\end{figure}

Fig.\ \ref{fig:sf} depicts the spectral function \eqref{eq:sf} for spin $\up$ in panel (a) and 
for spin $\down$ in panel (b)
for various values of $x$ near the Fermi energy $\omega=0$ with $\so=0.2t$, $S=1/2$, $\delta=7t$, and $S\jh=7.25t$.
For these values the system lies in the middle of the AFCI phase according to
Fig.\ \ref{fig:edelta}. 
For spin $\up$ in Fig.\ \ref{fig:sf}(a), there is clearly a finite spectral contribution at the Fermi energy
$\omega=0$ for $x=40$ indicating the existence of gapless excitations at the edge. The spectral contribution at $\omega=0$ 
disappears quickly as $x$ 
is moved away from the edge. The results for
$x=36$ already coincide with the results at $x=21$ which reproduce the bulk results
obtained for periodic boundary conditions in both directions. In stark contrast, 
the excitations are gapped in the bulk \emph{and} at the edges for spin
$\down$ in Fig.\ \ref{fig:sf}(b). This finding corroborates the existence of the AFCI further
because this analysis does not use the topological Hamiltonian.

\begin{figure}[t]
   \begin{center}
   \includegraphics[width=0.47\textwidth,angle=-90]{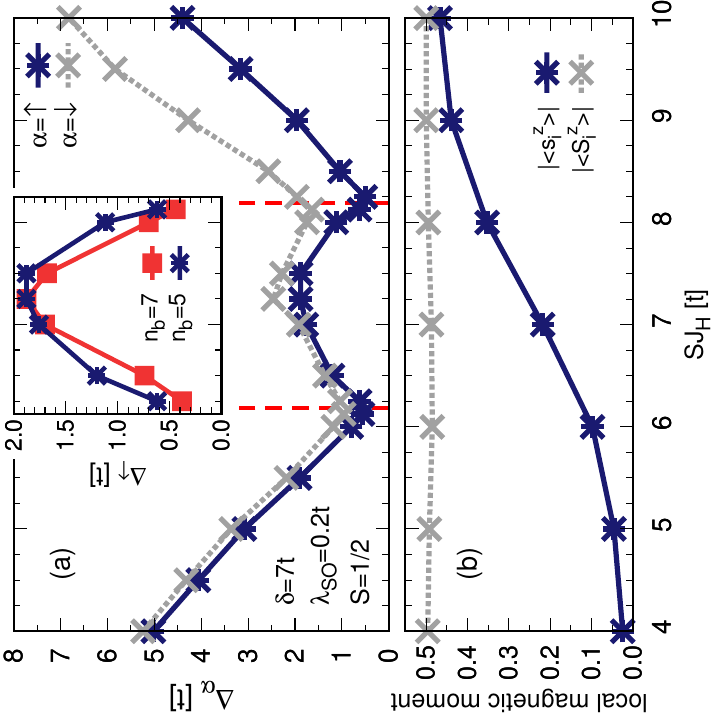}
   \caption{{\bf Charge gap closing across topological phase transitions.} (a) Charge gap in the bulk for the spin component $\alpha$ of the KMK model \eqref{eq:kmk}
   for the $z$-AF solution vs.\ the rescaled Hund coupling $S\jh$. Alternating sublattice potential
   $\delta=7t$, spin-orbit coupling $\so=0.2t$, and 
   localized spin $S=1/2$. The results are for the $z$-AF solution with the local magnetic moment
   $\langle s_i^z \rangle$ positive on the higher-energy sublattice $A$ and negative on the lower-energy
   sublattice $B$, which is why the spin $\up$ charge gap is smaller than the spin $\down$ charge gap.
   The vertical dashed lines denote the transitions predicted 
   by the analysis of the topological Hamiltonian. The data are computed of 
   $n_b=5$ bath sites in the ED impurity solver of the DMFT.  The inset compares the spin $\up$ 
	charge gap for $n_b=5$ and $7$ in the $z$-AFCI phase.
	(b) Magnitude of the local magnetic moment of itinerant electrons $|\langle s_i^z \rangle|$ and
	the localized spin $|\langle S_i^z \rangle|$ for the same model parameters as in panel (a).}
   \label{fig:gap}
   \end{center}
\end{figure}

The change of a Chern number in the transition from the trivial $z$-AF insulator
to the topological $z$-AFCI must be accompanied by the closing of the bulk charge gap
in the interacting model \eqref{eq:kmk}. Such gap closure is a necessary
prerequisite of this transition independent of the topological Hamiltonian
which is predicting the transition. Thus, we want to check whether
this necessary condition is fulfilled for the transitions predicted by the topological Hamiltonian
in Fig.\ \ref{fig:edelta}.  Fig.\ \ref{fig:gap}(a) displays the charge gap
$\Delta_\alpha$ for the spin $\alpha$ of the KMK model for the $z$-AF solution as function of
the rescaled Hund coupling $S\jh$. The model parameters are $\delta=7t$, $\so=0.2t$, and $S=1/2$.
Note Fig.\ \ref{fig:gap}(a) displays the results of the $z$-AF solution even for large $S\jh$
because here we study the charge gap closing across continuous topological phase transitions, ignoring
the first-order spin-flop transition.

The spin-resolved charge gap $\Delta_\alpha$ is given by the difference of the energy of the electron and the hole
peaks closest to the Fermi energy $\omega=0$ in the corresponding single-particle spectral function, 
see Figs.\ \ref{fig:sf}(a) and (b).
The vertical dashed lines in Fig.\ \ref{fig:gap}(a) specify the
transition points predicted by means of the topological Hamiltonian. The data in 
Fig.\ \ref{fig:gap}(a) clearly indicate minima of the charge gaps at these transitions.
The inset compares the spin $\up$ charge gap
for $n_b=5$ and $7$ in the region of the $z$-AFCI phase. 
In the middle of this region, both evaluations agree nicely while 
the minima at the borders decrease upon increasing the number of bath sites.
This is exactly what is to be expected because capturing a truly vanishing gap at the transition point,
i.e., a single metallic point, would
require a continuous representation of the bath, i.e., an infinite number of bath sites.
We conclude that the results in Fig.\ \ref{fig:gap}(a) is fully consistent with a vanishing spin $\up$ charge
gap at the dashed lines. This confirms that the topological phase transitions predicted by the topological Hamiltonian in Fig.\ \ref{fig:edelta} is indeed accompanied by the charge gap closing in the KMK model.

We point out that the energy difference between the spin $\up$ and the spin $\down$ charge gaps
in Fig.\ \ref{fig:gap}(a) is propotional to the magnitude of the local magnetic moment of the itinerant electrons 
$|\langle s_i^z \rangle|$. Fig.\ \ref{fig:gap}(b) depicts the magnitude of the local magnetic moment 
of itinerant electrons $|\langle s_i^z \rangle|$ as well as the localized spin $|\langle S_i^z \rangle|$ 
for the same model parameters as in  Fig.\ \ref{fig:gap}(a).
While $|\langle S_i^z \rangle|$ is almost always fully polarized close to $0.5$ ($\hbar=1$) the local magnetic moment
of itinerant electrons $|\langle s_i^z \rangle|$ is small for weak rescaled Hund coupling $S\jh$ and 
increases upon increasing $S\jh$. This explains the energy difference between the spin $\up$ and the spin $\down$ charge gaps in Fig.\ \ref{fig:gap}(a).

\section*{Discussion}
In this report, we provided four pieces of evidence for the generic phase diagram at zero temperature shown in
Fig.\ \ref{fig:main}(a). First, we considered the ground state energies of the phases under study
to determine which one represents the true state of the system. Second, we computed the Chern number in the bulk based on
the topological Hamiltonian. Third, we corroborated the topological character of the intermediate $z$-AFCI
phase by computing the gapless edge modes directly for the KMK model \eqref{eq:kmk}
in a stripe-like confined geometry. Finally, we found evidence that the charge gap in the KMK model
vanishes at the predicted transitions by the topological Hamiltonian between topologically trivial
and non-trival phases. Furthermore, we uncovered in Fig.\ \ref{fig:main}(c) that the charge gap in the
$z$-AFCI phase exhibits a remarkable increase upon increasing the spin-orbit coupling. In most experimental
realization, $\so \leq 0.15t$ and hence the charge gap in the $z$-AFCI is very well described by
the ready-to-use formula \eqref{eq:gap}.

What is the general relevance of the obtained findings beyond the particular model studied?
Although our analysis is performed for the honeycomb lattice we expect that the main findings 
on the existence of the collinear AFCI and the charge gap equation \eqref{eq:gap} are generic 
and apply also for other similar bipartite lattices. Indeed, it has already been shown that the
existence of the collinear AFCI is not restricted to the honeycomb lattice, but can also be found
in other types of lattices \cite{Ebrahimkhas2022}.
The phase diagram obtained for the extended time-reversal-invariant Harper-Hofstadter-Hubbard model defined
on the square lattice \cite{Ebrahimkhas2022} indicates close similarity to the phase diagram obtained for
the Kane-Mele-Hubbard model defined on the honeycomb structure \cite{Jiang2018}. This suggests that
the AFCI is a generic feature of systems with strong spin-orbit coupling and electronic correlation and
exists independent of the lattice structure and the details of the model.
The essential prerequisite is
the absence of a point group operation which compensates the effect of the time-reversal transformation
on the electronic state \cite{Ebrahimkhas2022}.

In the Kane-Mele-Kondo model, no Hubbard repulsion $U$ appears. Including this interaction
is beyond the scope of the present work, but can also be done within the DMFT treatment.
Based on previous results, however, we do  not expect qualitative changes in the physical behavior
since the effect of the Hubbard and the Hund coupling reinforce each other in suppressing double occupancies
 \cite{Dagotto1998,MuellerHartmann1996}. The gap between a single and double occupancy on an isolated site
change from $S\jh$ to $U/2+S\jh$. According to this observation,
the inclusion of the Hubbard interaction is likely to shift the phase boundaries
and we conjecture that the AFCI will appear 
at $\delta \approx U/2 + S\jh$ rather than at $\delta \approx S\jh$.
The Rashba spin-orbit coupling neglected in Eq. \eqref{eq:kmk} is also present in real materials.
But it is rather weak and thus it is not expected to be able to destroy the AFCI phase;
neither is it likely to have a significant effect on the charge gap \cite{Liu2011}.
Nevertheless, the study of the effect of the Rashba spin-orbit coupling on AF topological
insulators deserves future attention.

Let us discuss in which experimental systems an AFCI is likely to appear.
The prediction of the AFCI and of the magnitude of its charge gap in different transition-metal
compounds requires quantitative knowledge of the spin-orbit coupling. This is a challenging task. 
Using the average values of $4.5$, $15$, and $40$ meV
respectively for $3d$, $4d$, and $5d$ transition-metal elements \cite{Khomskii2021} one can at least
estimate orders of magnitude. Note that these numbers refer to the Kane-Mele-type of the spin-orbit coupling
$\so$ in honeycomb lattices as defined in Eq.\ \eqref{eq:kmk}. A precise value for $\so$ in silicene
is $0.77$ meV, in germanene  $8.9$ meV, and in stanene $12$ meV \cite{Liu2011}.
For $3d$ materials, one finds that a hopping amplitude 
$t\lesssim \so/0.03= 150$ meV is necessary to stabilize the AFCI.
Although this appears small compared to the hopping amplitude $370$ meV estimated for
cuprates \cite{Coldea2001a,Sheshadri2023}, it can still be realized in some $3d$ compounds.
Monolayers of CrO are reported to show collinear antiferromagnetic ordering with a
high N\'eel temperature $T_{\rm N}\approx 750$ K \cite{Chen2023}. The antiferromagnetic NN
exchange interaction $J$ is by far the dominant term with  $J \approx 10$ meV for the spin quantum number 
$\mathcal{S}=2$ of  Cr$^{+2}$ ions \cite{Chen2023}.
Reasonable estimates of the Hubbard interaction $U=4.5$ eV and the Hund coupling $\jh=1$ eV
for Cr imply an effective interaction $U_{\rm eff}=U-\jh=3.5$ eV agreeing with the findings of
Ref.\ \onlinecite{Chen2023}. 

From the exchange interaction $J={4t^2}/{(U+2S\jh)}$
where $U+2S\jh$ is the bare Mott gap we deduce the estimate for the hopping parameter $t\approx 140$ meV.
Here, we use $S=\mathcal{S}-1/2=3/2$ for the localized spin in a Kondo lattice while
the deducted $1/2$ refers to the spin of the itinerant electrons \cite{Hafez-Torbati2021,Hafez-Torbati2022}. 
This implies that a monolayer of CrO is a potential candidate
to realize the AFCI with a charge gap $\Delta \approx 5$ meV based on Eq.\ \eqref{eq:gap}. 
This prediction agrees roughly with the finding of an AFCI with a charge gap of about 1 meV 
in a first-principle calculations \cite{Guo2023}
for monolayers of CrO subjected to biaxial and shear strains.

In general, larger hopping elements are expected for $4d$ materials because
$4d$ orbitals are more extended than $3d$ orbitals. Assuming a range of hoppings of $300$-$500$ meV for $4d$ materials one is led to predict the possible AFCIs 
 with charge gaps ranging from $0$ to $80$ meV, or in other words, with charge gaps of
the order of a few tens of meV. 
This is to be compared with the charge gap of a few meV we estimated above for the AFCI in $3d$ materials.
A monolayer of MoO is an example of a $4d$ transition-metal oxide which is predicted to host
 collinear AF ordering with a N\'eel temperature as high as $T_{\rm N} \approx 1100$ K
 based on first-principle calculations \cite{Wu2023}. Subjected to
strain, a collinear AFCI with a charge gap of about $50$ meV is expected \cite{Wu2023}. This agrees  with our estimate for
the charge gap of AFCI in $4d$ materials. The proposed effective Heisenberg spin model 
for MoO includes a non-negligible ferromagnetic next-nearest-neighbor  interaction in addition to the
antiferromagnetic nearest-neighbor interaction \cite{Wu2023}. Unfortunately, this prevents to employ
 Eq.\ \eqref{eq:gap} for a more accurate comparison.

It should be mentioned that
the lattice structure of CrO and MoO are different from the honeycomb lattice and a quantitative
description of these compounds may require a model with more than a single itinerant orbital.
However, we are confident that these details will not lead to an order of magnitude change in the charge gap
and that the KMK model in Eq. \eqref{eq:kmk} can still capture qualitatively the main physics.

For the even larger range of 600-1000 meV for the hopping elements for $5d$ compounds we deduce a
charge gap $\Delta \approx 140$-310 meV. Finally, bilayers of perovskite-type made
from heavy transition-metal oxides grown along the $[111]$ crystallographic axis also exhibit
a buckled honeycomb structure. The alternating sublattice potential can be fine-tuned by a
perpendicular electric field or by placing different substrates on the top and the bottom layers \cite{Xiao2011}.
In summary, our results and estimates provide evidence for the possibility to realize
antiferromagnetic Chern insulators in $4d$ and $5d$ transition-metal compounds
persisting up to the room temperatures.

We emphasize that a large-gap AFCI at room temperature would be highly attractive for future
fundamental and applied research. Samples of AFCI would render investigations of robust 
topological phenomena possible such as the unidirectional transport of charge and 
spin properties along the edge modes.
Thereby, one further important step towards the vision of fast information processing with 
little dissipation based on antiferromagnetic magnons will be made.
This would open up a plethora of conceptual and applied research directions.

\section*{method}

{\small
{\bf Dynamical mean-field theory.} To address the KMK model in the range 
from weak to strong Hund coupling we employ the non-perturbative method of DMFT
mapping the lattice model \eqref{eq:kmk} to an effective Kondo impurity
problem \cite{Hafez-Torbati2021,Hafez-Torbati2022}.
The approach fully takes into account all local quantum fluctuations, but neglects non-local ones 
by approximating the self-energy to be local in space,
$\Sigma_{ij;\alpha\beta}({\rm i}\omega_n)=\Sigma_{i;\alpha\beta}({\rm i}\omega_n)\delta_{ij}$
where $i,j$ are site indices and $\alpha, \beta\in\{ \up,\down\}$ are spin indices.
The non-local elements of the self-energy could be included in 
cluster extensions of the DMFT \cite{Park2008,Potthoff2018}.
The method is already employed to study
the interaction driven topological phase transitions in various models \cite{Yu2011,Wu2012,Wu2016,Gu2019}.
The cluster DMFT has the disadvantage of breaking the translational symmetry of the lattice.
Yet this can be cured by periodization procedures  which, however, can lead to spurious
nonzero Chern numbers \cite{Gu2019}. Hence, we focus on the single-site DMFT
in this study.
Non-zero off-diagonal elements in the spin quantum numbers, i.e.,  $\alpha \neq \beta$, are indispensable to find
$xy$-AF solutions \cite{Hafez-Torbati2018}. We use the real-space realization of the DMFT which enables us to
investigate bulk as well as edge properties on equal footing \cite{Potthoff1999,Song2008,Snoek2008}.
We specifically use the implementation
discussed in Ref.\ \cite{Hafez-Torbati2018}. We treat the honeycomb lattice as a brick wall 
labeling each lattice site
with two integers $x$ and $y$, see Fig.\ \ref{fig:lattice}. We consider lattices of
the size $L_x \times L_y=40 \times 40$ with periodic boundary conditions in both directions to
explore bulk properties. Near the critical regions with small or vanishing gaps 
we produced data also for $L_x \times L_y=60 \times 60$ to make sure that the results are independent 
of the lattice size.
To address edge modes we consider cylindrical geometries with
the open boundary condition in  $x$ direction and periodic ones in $y$ direction 
for $L_x \times L_y=41 \times 40$.

\begin{figure}[t]
   \begin{center}
   \includegraphics[width=0.44\textwidth,angle=0]{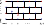}
   \caption{{\bf Honeycomb lattice as a brick wall.} Honeycomb lattice as a brick wall where the sites are labeled by
   two integers $(x,y)$. The number of lattice sites along $x$ is $L_x$ and along $y$ is $L_y$.
   Sublattices $A$ and $B$ are specified. We use periodic boundary conditions in both directions
   to study bulk properties and periodic boundary condition in $y$ and open boundary condition
   in $x$ direction to study edge states.}
   \label{fig:lattice}
   \end{center}
\end{figure}

We employ exact diagonalization \cite{Caffarel1994} to solve the Kondo impurity model
because it allows us to treat the localized spin without
any truncation in contrast to the quantum Monte Carlo (QMC) solver which suffers from the 
notorious fermionic sign problem \cite{Yunoki1998,Peters2006} if the spin flip
terms of the Kondo interaction are included.
In addition, the ED provides direct access to the real-frequency spectral functions in contrast to
the QMC which requires analytical continuation.
Due to the finite number of bath sites in the ED the spectral function consists of
sharp peaks approximating a continuous function. However, the charge gap extracted from the
spectral function is found to be accurate; the ED results are used to benchmark the QMC results 
\cite{Wang2009}. We provide data for various numbers of bath sites, especially near the
critical regions, to show that the obtained results do hardly depend on the number
of bath sites.

{\bf Topological Hamiltonian.}
We use the topological Hamiltonian $H_{\rm top}$ to compute the topological invariants \cite{Wang2012}.
The method  \cite{Wang2013} relates the topological invariant of an interacting
fermionic system to an effective non-interacting model given by the  topological Hamiltonian which  reads
in second quantization
\be
\label{eq:th}
H_{\rm top}=H_{0}+\sum_{ij}\sum_{\alpha\beta} c^{\dag}_{i\alpha} \Sigma^{\vpdag}_{ij;\alpha\beta}(0) c^{\vpdag}_{j\beta} \ ,
\ee
where $H_{0}$ is the non-interacting, bilinear part of the original Hamiltonian and 
$\Sigma_{ij;\alpha\beta}(0)$ is the self-energy at zero frequency. The method relies on the 
adiabatic deformation of the imaginary-frequency
Green function such that the charge gap never closes \cite{Wang2012}.

The second term in Eq.\ \eqref{eq:th} becomes local in the DMFT approximation. In addition, in the $z$-AF phase
the self-energy has no off-diagonal elements in spin space leading to 
$\Sigma_{i;\alpha\beta}(0)=\Sigma_{i;\alpha}(0)\delta_{\alpha\beta}$.
The symmetry $\Sigma_{A;\alpha}(0)=-\Sigma_{B;\alpha}(0)\in\mathds{R}$ between the zero-frequency self-energies
on sublattices $A$ and $B$ simplifies the topological Hamiltonian in the $z$-AF phase to the Kane-Mele model 
with an \emph{effective} spin-dependent sublattice potential given in Eq.\ \eqref{eq:edelta}.

In the $xy$-AF phase the spin diagonal elements obey the spin symmetry
$\Sigma_{i;\uparrow\uparrow}(0)=\Sigma_{i;\downarrow\downarrow}(0)=:\Sigma_{i}(0)$ and the 
 symmetry $\Sigma_{A}(0)=-\Sigma_{B}(0)$ between the two sublattices. 
For the off-diagonal elements in spin space we have
$\Sigma_{i;\uparrow\downarrow}(0)=\Sigma^*_{i;\downarrow\uparrow}(0)$ and
$\Sigma_{A;\uparrow\downarrow}(0)=-\Sigma_{B;\uparrow\downarrow}(0)$.
As a result the topological Hamiltonian \eqref{eq:th} in the $xy$-AF phase is described
by the Kane-Mele Hamiltonian with an effective spin-independent sublattice potential
$\tilde{\delta}=\delta+\Sigma_{A}(0)$ subjected to the alternating in-plane magnetic field
\be
\label{eq:eh}
\vec{h}_{A}=-\vec{h}_{B}=2(-{\rm Re}[\Sigma_{A;\uparrow\downarrow}(0)],
~+{\rm Im}[\Sigma_{A;\uparrow\downarrow}(0)],~0)^\top \ ,
\ee
where $\vec{h}_{A}$ and $\vec{h}_{B}$ act on  sublattice $A$ and $B$, respectively.
The term ``in-plane'' refers to the $xy$ plane.
Similar to the uniform in-plane magnetic field in Ref.\ \onlinecite{Kane2005b} 
even an infinitesimal alternating in-plane magnetic
field \eqref{eq:eh} gaps the edge modes and the system becomes a trivial band insulator. 
Hence, the $xy$-AF phase is always topologically trivial.
We recall that even when $S_z$ is not conserved the QSHI and the CI can still exist
and the topological invariants can still be defined \cite{Kane2005b,Yang2011a}. In our case,
however, either $S_z$ is conserved or the system is topologically trivial.

The combination of the DMFT and the approach based on the 
topological Hamiltonian has been extensively used to map out
the phase diagram of various interacting topological models
of fermionic character
\cite{Hofstetter2018,Budich2013,Amaricci2015,Irsigler2019,Hafez-Torbati2020,Ebrahimkhas2021,Ebrahimkhas2022,Vanhala2016}.
The phase diagram of the Haldane-Hubbard model obtained using the DMFT and the topological Hamiltonian 
agrees qualitatively with the results obtained using ED for finite clusters and twisted boundary conditions 
\cite{Vanhala2016}.
A systematic study of non-local contributions to the self-energy also indicates only small changes to the phase
boundaries \cite{Mertz2019}
supporting to use the DMFT approximation of a local self-energy. Thus, the presented results 
are expected to be qualitatively reliable and that the inclusion of non-local
quantum fluctuations only leads to some quantitative corrections.
}

\section*{Data availability}
Data supporting the findings of this study are available within the article.


%

\begin{acknowledgments}
We would like to thank Elham Bayatloo for implementing a C++ class for the honeycomb lattice so that
the Kane-Mele-Kondo model could be easily implemented in the existing real-space DMFT code.
\end{acknowledgments}

\end{document}